# The effect of induced flow on the water wave skewness

Alexey V. Slunyaev[1,2,a)] and Anna V. Kokorina[2]

[1)] National Research University Higher School of Economics, Nizhny Novgorod, Russia
[2)] Institute of Applied Physics of the Russian Academy of Sciences, Nizhny Novgorod, Russia

[a)] Author to whom correspondence should be addressed: slunyaev@ipfran.ru

**Abstract**
Nonlinear wave-induced perturbations are discussed within the framework of the second-order theory. Due to the slow attenuation of the long perturbations with depth, they modulate motions beneath surface waves down to the bottom and can strongly affect the wave statistics. The disturbance which corresponds to the second harmonic, at a sufficient depth inverses the phase, thereby qualitatively changes the shape of individual waves. The skewness coefficients of the surface displacement and the bottom pressure fluctuations are considered. It is shown that the bottom pressure skewness may have different signs and magnitudes depending on the water depth and the configuration of nonlinear wave groups which compose the wave field. The effect on the surface displacement skewness is much weaker. Mechanisms leading to strong deviation of the skewness coefficient from the nonlinear Stokes wave theory are discussed.

## 1. Introduction

It is traditionally believed that wind waves on the sea surface have positive values of the skewness coefficient, which is the third statistical moment characterizing the vertical asymmetry of waves. The theoretical justification of this point is presented in Fig. 1, where the skewness $Sk^{(\eta)}$ of the surface displacement normalized by the wave steepness is plotted for the example of a regular nonlinear wave computed as the Stokes wave solution in different depths. In the figure, the wave height is taken equal to 40% of the breaking limit for the given depth. The ratio of skewness over wave steepness tends to a positive constant in deep water and grows as the water becomes shallower. The curves corresponding to the 2-order and 5-order Stokes solutions [Fenton, 1990] agree very well (circles and solid line in Fig. 1 respectively) confirming that waves are skewed mainly due to the second harmonic and that the second-order nonlinear Stokes theory serves well in the range of deep and intermediately deep water when waves are moderately steep. In shallower water waves become cnoidal, and the positive vertical asymmetry of waves with sharper crests and flatter troughs becomes even stronger. The theoretical formula for the skewness of deep-water random Gaussian waves represents a linear relation to the wave steepness, $Sk^{(\eta)} = 3k_0\eta_{\mathrm{rms}}$, where $k_0$ is the characteristic wavenumber and $\eta_{\mathrm{rms}}$ is the root-mean-square surface displacement [Mori & Janssen, 2006]. Thus, the skeweness of irregular nonlinear deep-water waves should be positive too.

Meanwhile the skewness coefficients measured in-situ exhibit remarkably large spread of values, see e.g. in [Soares et al., 2004; Zapevalov & Garmashov, 2022; Zapevalov, 2025] and references therein. Besides showing rather uncertain dependence on the nonlinearity parameter (i.e. characteristic wave steepness), the measured skewness of surface displacements can reach essentially negative values what is qualitatively different from the generally accepted picture outlined above.

The results of in-situ measurements of the bottom pressure variations off the coast of Sakhalin Island presented in Fig. 2 reveal an essentially richer picture of the skewness coefficient $Sk^{(p)}$ calculated for the bottom pressure fluctuations. The data were collected during several experimental campaigns in 2012-2022 by autonomous sensors installed on the flat sea bed at the depth about 10 m with the acquisition frequency 1–8 Hz, see details in [Slunyaev et al., 2023]. The recorded long series of data were divided into 20-minute samples after filtering out oscillations longer than 10 minutes. For each sample, the parameter $k_0h$ (where $k_0$ is the wavenumber calculated from the mean zero-crossing period according to the linear dispersion law, and $h$ is the depth) and the skewness $Sk^{(p)}$ were computed, which correspond to the points in Fig. 2. The concentric contours show the data density, so that the greatest number of data corresponds to the center. The significant range of dimensionless depths $k_0h$ in the figure is due to the effect of tides and varying wave period. On the one hand, the data demonstrates a large scatter of the skewness values which may be of different signs; on the other hand the data exhibits clear dependence on the depth parameter.

Negative skewness of pressure fluctuations were registered at different depths beneath intense irregular waves measured in laboratory conditions [Slunyaev et al., 2022]. The experiment was performed in deep water, $k_0h = 3$, while the pressure was registered at different depths from $k_0h = 0.37$ to $k_0h = 1.65$. The skewness of pressure was negative at any depth gradually reducing in the absolute value from about –0.4 just below the wave troughs.

In this paper we analyze theoretically the skewness coefficients of the surface displacement and of the bottom pressure of finite-depth water waves. We show that in addition to the nonlinear second harmonic, the skewness coefficient is affected by the wave-induced long-scale perturbations. This effect is particularly important beneath the modulated nonlinear waves in intermediate depth $k_0h = O(1)$ since for these conditions the induced perturbations are strong and attenuate with depth slowly. Such perturbations formally correspond to the difference (or zeroth) nonlinear wave component. Their importance for marine engineering was emphasized in e.g. [Dogliani & Cazzulo, 1992], but despite this fact they are still ignored. The present investigation is partly related to the work by [Chen, 2006] where the Stokes wave set-down was considered as a limit of bi-chromatic waves. Obtained there 2-nd order corrections to the classic solution were used in a later study [Rijnsdorp, 2022].

In the present work we show that the induced long-scale movements under appropriate conditions can lead to negative values of the bottom pressure skewness in large and intermediate depths. Waves of different group structures can arise, what radically changes the overall picture of irregular wave asymmetry. This issue has not been discussed in the literature before.

The paper is organized as follows. The second-order nonlinear theory for the bottom pressure beneath water waves traveling in intermediate depth is revisited in Sec. 2 with a particular focus on the wave-induced long-scale perturbations. The process of the long-scale bound wave generation is considered in Sec. 3 using the direct numerical simulation of hydrodynamic equations. The effect of the second-order nonlinear bound waves on the statistical moments is examined in Sec. 4 for the general setting of the problem; there the universal expressions for the skewness coefficient are obtained. The skewness coefficients for the hydrodynamic fields of the surface displacement and the bottom pressure are obtained and discussed in Sec. 5. It is shown that the skewness value is highly sensitive to the background wave features. Relation of the theoretical results to the in-situ data of the bottom pressure measurements and also stochastic numerical simulations is discussed in Sec. 6. It is emphasized that the skewness strongly depends on the assumed structure of wave groups which compose the stochastic wave field. Three basic group structures are suggested which may occur depending on the circumstances. Conclusions are collected in Sec. 7.

## 2. Second-order nonlinear hydrodynamic theory for narrowband waves

### Governing equations and short-wave components

Let us consider the gravity water wave problem in planar configuration with $Ox$ being the horizontal axis directed along the wave propagation and $Oz$ the upward axis. The undisturbed water level corresponds to $z = 0$, while the flat bottom is located at $z = -h$, where the constant $h$ will be referred to as the water depth. The water is assumed to be uniform and ideal under the action of the gravity force with the acceleration $g = 9.81$ m/s$^2$. The disturbed water surface is prescribed by the function $\eta(x,t)$. The water movement will be assumed potential with the velocity potential $\varphi(x,z,t)$ which satisfies the Laplace equation in the water column

$$\frac{\partial^2 \varphi}{\partial x^2} + \frac{\partial^2 \varphi}{\partial z^2} = 0, \quad -h \leq z \leq \eta. \tag{1}$$

The bottom boundary condition is non-leaking,

$$\frac{\partial \varphi}{\partial z} = 0, \quad z = -h. \tag{2}$$

The surface boundary conditions consist of the kinematic and dynamic equations respectively:

$$\frac{\partial \eta}{\partial t} - \frac{\partial \varphi}{\partial z} = -\frac{\partial \eta}{\partial x}\frac{\partial \varphi}{\partial x}, \quad z = \eta, \tag{3}$$

$$g\eta + \frac{\partial \varphi}{\partial t} = -\frac{1}{2}\left(\frac{\partial \varphi}{\partial x}\right)^2 - \frac{1}{2}\left(\frac{\partial \varphi}{\partial z}\right)^2, \quad z = \eta. \tag{4}$$

Equations (1)-(4) represent the standard set of hydrodynamics equations sufficient for description of many problems of water wave dynamics, see e.g. [Johnson, 1997; Mei et al., 2005]. Key notations adopted in this work are collected in Table 1.

It is well known that the general solution of the Laplace equation (1) with the bottom boundary condition (2) may be presented in the form

$$\varphi(x,z,t) = \int_{-\infty}^{\infty} C_k(t)\Psi_k(z)\exp(-ikx)dk, \quad \Psi_k(z) = \frac{\cosh(k(z+h))}{\cosh kh}, \tag{5}$$

$$C_k(t) = \frac{1}{2\pi}\int_{-\infty}^{\infty} \varphi(x,0,t)\exp(ikx)dx.$$

Here $C_k(t)$ are the Fourier coefficients and $\Psi_k(z)$ are corresponding vertical modes of the solution; the spectral variable $k$ has the meaning of the wavenumber.

The velocity potential in close vicinity of $z = 0$ may be decomposed using the Taylor series:

$$\varphi(x,z,t) \approx \varphi(x,0,t) + z\frac{\partial \varphi}{\partial z}\bigg|_{z=0}. \tag{6}$$

Under the assumption that the wave amplitude is small, this expansion may be used to relate the potential on the surface $z = \eta$ with the potential at the rest water level $z = 0$. The derivatives of $\varphi$ with respect to $x$, $z$ and $t$ can be calculated on the surface similarly. In this work, these expansions are limited to accounting of up to quadratic nonlinear terms.

In what follows we develop the second-order nonlinear theory for narrow-banded waves assuming that the solution is well represented by the dominant harmonic with a given wavenumber $k_0 > 0$, and two components generated by the 'plus' and 'minus' nonlinear interactions. The second harmonic corresponds to wavenumbers about $2k_0$, while the difference (or zeroth) harmonic is characterized by small wavenumbers, $k/k_0 \ll 1$:

$$\eta = \eta^{(0)}(x,t) + \eta^{(I)}(x,t) + \eta^{(II)}(x,t), \quad \varphi = \varphi^{(0)}(x,z,t) + \varphi^{(I)}(x,z,t) + \varphi^{(II)}(x,z,t), \quad (7)$$

$$\eta^{(I)} = \frac{1}{2}\left(A(x,t)e^{i\theta} + c.c.\right), \quad \eta^{(II)} = \frac{1}{2}\left(A_2(x,t)e^{2i\theta} + c.c.\right),$$

$$\varphi^{(I)} = \frac{1}{2}\left(B(x,t)e^{i\theta} + c.c.\right)\Psi_{k_0}(z), \quad \varphi^{(II)} = \frac{1}{2}\left(B_2(x,t)e^{2i\theta} + c.c.\right)\Psi_{2k_0}(z),$$

$$\theta(x,t) = \omega_0 t - k_0 x.$$

Here the upper indices denote the harmonics; $A(x,t)$ and $B(x,t)$ are complex envelopes of the carrier wave; $A_2$ and $B_2$ are complex amplitudes which represent the second wave harmonic; the functions $\eta^{(0)} = <\eta>_T$ and $\varphi^{(0)} = <\varphi>_T$ describe the long-scale surface displacement and flow, respectively. By $<\cdot>_T$ we denote averaging over a wave period for a given coordinate. All the functions except the phase $\theta$ depend on the variables $x$ and $t$ slowly.

As the problem setup is classic, we will take the formulas describing the dominant wave and the second harmonic from [Slunyaev et al., 2022] without derivation. Though the theory for the difference harmonic will be reproduced below in more detail. The ansatz (7) should be substituted to the governing equations (3) and (4) using the Taylor expansion (6) for the velocity potential and its derivatives. In the first approximation the terms of the second and zeroth harmonics are disregarded, and the dispersion relation between the carrier wavenumber $k_0$ and the carrier frequency $\omega_0$ appears as a compatibility condition of the system of two equations on the amplitudes $A$ and $B$,

$$\omega_0 = \sqrt{gk_0\sigma}, \quad \sigma = \tanh(k_0 h). \quad (8)$$

If the dispersion relation (8) holds, the amplitudes are related as follows

$$B = i\frac{g}{\omega_0}A. \quad (9)$$

The nonlinear corrections appear in the next order. Assuming that the harmonics overlap weakly, the terms which correspond to different harmonics may be treated independently. In particular, the amplitudes of the second harmonic are fully determined by the carrier wave amplitude (see Eqs. (42) and (45) in [Slunyaev et al., 2022]):

$$A_2 = k_0 \frac{3-\sigma^2}{4\sigma^3} A^2, \quad B_2 = 3i\omega_0 \frac{1-\sigma^2}{8\sigma^4} A^2. \quad (10)$$

**The long-scale component**

For the long-scale part of the solution the system of equations resulting from Eqs. (3) and (4) is the following:

$$\frac{\partial \eta^{(0)}}{\partial t} - \left.\frac{\partial \varphi^{(0)}}{\partial z}\right|_{z=0} = -\frac{\omega_0}{2\sigma}\frac{\partial}{\partial x}|A|^2, \quad (11)$$

$$g\eta^{(0)} + \frac{\partial \varphi^{(0)}}{\partial t}\bigg|_{z=0} = -\frac{\omega_0^2}{4}\frac{1-\sigma^2}{\sigma^2}|A|^2. \quad (12)$$

It is assumed that the phase-locked long wave component convoys the 'mother' wave train with its velocity, which is the group velocity $C_{gr}$, and then

$$\frac{\partial \varphi^{(0)}}{\partial t}\bigg|_{z=0} = -C_{gr}\frac{\partial \varphi^{(0)}}{\partial x}\bigg|_{z=0}, \qquad \frac{\partial \eta^{(0)}}{\partial t} = -C_{gr}\frac{\partial \eta^{(0)}}{\partial x}, \quad (13)$$

where

$$C_{gr} = \frac{d\omega}{dk}\bigg|_{k_0} = \frac{C_{ph}}{2}\left(1 + k_0 h\frac{1-\sigma^2}{\sigma}\right), \qquad C_{ph} = \frac{\omega_0}{k_0}, \quad (14)$$

$C_{ph}$ is the phase velocity of the carrier wave. With the use of relation (13) the equation (12) may be written as follows

$$g\eta^{(0)} - C_{gr}\frac{\partial \varphi^{(0)}}{\partial x}\bigg|_{z=0} = -\frac{\omega_0^2}{4}\frac{1-\sigma^2}{\sigma^2}|A|^2. \quad (15)$$

The combination of equations (11) and (15) gives a close equation on $\varphi^{(0)}$ at the horizon $z = 0$ (which is equivalent to Eq. (49) in [Slunyaev et al., 2022]):

$$\left(g\frac{\partial \varphi^{(0)}}{\partial z} + C_{gr}^2\frac{\partial^2 \varphi^{(0)}}{\partial x^2}\right)\bigg|_{z=0} = \frac{g^2}{2C_{ph}}\left(1 + \frac{1-\sigma^2}{2}\frac{C_{gr}}{C_{ph}}\right)\frac{\partial |A|^2}{\partial x}. \quad (16)$$

The Laplace equation on $\varphi^{(0)}$ in the domain $-h \leq z \leq 0$ and the bottom boundary condition at $z = -h$ have the forms identical to Eqs. (1) and (2) respectively, and then the problem on the potential of the long-scale part of waves is fully determined. When $\varphi^{(0)}(x,z,t)$ for the given $A(x,t)$ is obtained, the solution for $\eta^{(0)}(x,t)$ follows from Eq. (15).

It is convenient to represent Eq. (16) in terms of the spatial Fourier transform

$$k^2\left[C_{ph}^2(k) - C_{gr}^2(k_0)\right]\hat{F}\{\varphi^{(0)}|_{z=0}\} = -ik\frac{g^2}{2C_{ph}}\left(1 + \frac{1-\sigma^2}{2}\frac{C_{gr}}{C_{ph}}\right)\hat{F}\{|A|^2\}, \quad (17)$$

where $\hat{F}\{\cdot\}$ denotes the spatial Fourier transformation and $k$ is the spectral variable. Note that the phase velocity $C_{ph}(k) = \omega(k)/k$ from the left hand side is a function of the spectral variable with $\omega(k)$ denoting the dispersion relation as per Eq. (8). The group velocity $C_{gr}(k_0)$ from the left-hand-side, and also $C_{ph}$ and $C_{gr}$ to the right are all calculated for the carrier wave, $k = k_0$.

The problem (17) can be solved numerically. The solution contains singularity when the expression in square brackets tends to zero, what corresponds to the synchronism between the induced long-scale flow and linear waves. This condition reads $C_{ph}(k) = C_{gr}(k_0)$ and can be easily analyzed graphically in the plane $(k,\omega)$. Its analytic representation leads to the transcendental equation

$$C_{ph}(k) = C_{ph}(k_0)\left(\frac{1}{2} + \frac{k_0 h}{\sinh(2k_0 h)}\right). \quad (18)$$

Under the conditions of deep water $k_0 h \gg 1$ the solution of Eq. (18) is $k \approx 4k_0$. The small-amplitude wave generation by a nonlinear wave group due to this mechanism was observed in numerical simulations by [Slunyaev, 2018]. In the shallow water limit $k_0 h < kh \ll 1$ the condition (18) is satisfied for significantly shorter waves, $k \approx 3^{1/2} k_0$.

Due to the narrowband assumption, the Fourier amplitudes of the function $|A|^2$ from the right-hand-side of Eq. (17) and consequently of the function $\varphi^{(0)}$, may have significant values at small wavenumbers only, therefore let us regularize the solution (17) by using the shallow-water limit for $C_{ph}(k)$: $C_{ph}(k) \approx C_{LW}$, where $C_{LW} = (gh)^{1/2}$ is the long-wave velocity, $C_{LW} \geq C_{ph}(k) \geq C_{gr}(k)$. Then from Eq. (17) we obtain expression for the long-scale horizontal flow calculated at the horizon $z = 0$ in the form:

$$U(x,t) = \frac{\partial \varphi^{(0)}}{\partial x}\bigg|_{z=0} = \frac{-g^2}{2C_{ph}(C_{LW}^2 - C_{gr}^2)}\left(1 + \frac{1-\sigma^2}{2}\frac{C_{gr}}{C_{ph}}\right)|A|^2 + U_\infty. \quad (19)$$

Here $U_\infty$ is the integration constant which has the meaning of the background horizontal velocity in the area where the wave is absent (i.e., $A = 0$).

With the same degree of accuracy, the factor in the left-hand-side of Eq. (17) may be presented in a different form:

$$k^2\left(C_{ph}^2(k) - C_{gr}^2(k_0)\right) \approx k^2 C_{ph}^2(k)\left(1 - \frac{C_{gr}^2(k_0)}{C_{LW}^2}\right), \quad (20)$$

and then Eq. (17) yields

$$\frac{\partial \varphi^{(0)}}{\partial z}\bigg|_{z=0} = \frac{gC_{LW}^2}{2C_{ph}(C_{LW}^2 - C_{gr}^2)}\left(1 + \frac{1-\sigma^2}{2}\frac{C_{gr}}{C_{ph}}\right)\frac{\partial |A|^2}{\partial x}. \quad (21)$$

Using relation (19), from Eq. (15) we obtain the expression for the long-scale displacement

$$\eta^{(0)} = -\frac{k_0}{4\sigma}\frac{2C_{ph}C_{gr} + C_{LW}^2(1-\sigma^2)}{C_{LW}^2 - C_{gr}^2}|A|^2 + \eta_\infty, \quad (22)$$

where $\eta_\infty$ is the integration constant, which may be referred to as the outer reference level, $\eta^{(0)} = \eta_\infty$ where $A = 0$. According to Eq. (15), the two constants of integration are related: $g\eta_\infty = C_{gr}U_\infty$.

If one puts $\eta_\infty = 0$, the obtained solutions (19), (21) and (22) correspond to the classic nonlinear Schrödinger theory for weakly modulated slowly modulated waves in finite depth water, see e.g. the book [Johnson, 1997].

**Bottom pressure**

Let us now calculate the hydrodynamic pressure at the bottom for the three main wave harmonics taking into account quadratic nonlinear terms. According to the Bernoulli law, the normalized by the water density $\rho$ dynamic pressure $P(x,z,t)$ is defined by the formula

$$P = \frac{1}{\rho}P_{tot} + gz = -\frac{\partial \varphi}{\partial t} - \frac{1}{2}\left(\left(\frac{\partial \varphi}{\partial x}\right)^2 + \left(\frac{\partial \varphi}{\partial z}\right)^2\right), \quad (23)$$

where $P_{tot}$ is the total pressure. When Eq. (23) is written for the bottom pressure $P_b(x,t) = P(x,z=-h,t)$, the last term vanishes due to the boundary condition (2), and then

$$P_b = P(x, z = -h, t) = -\frac{\partial \varphi}{\partial t}\bigg|_{z=-h} - \frac{1}{2}\left(\frac{\partial \varphi}{\partial x}\bigg|_{z=-h}\right)^2. \quad (24)$$

For the wave represented by three harmonics as per Eq. (7), the corresponding expression for the bottom pressure reads

$$P_b(x,t) = p_b^{(0)}(x,t) + p_b^{(I)}(x,t) + p_b^{(II)}(x,t), \qquad (25)$$

where $p_b^{(I)} = \frac{1}{2}\left(p(x,t)e^{i\theta} + c.c.\right)$ and $p_b^{(II)} = \frac{1}{2}\left(p_2(x,t)e^{2i\theta} + c.c.\right)$.

The expression for the first harmonic appears straightforwardly in the linear theory:

$$p_b^{(I)} = -\left.\frac{\partial \varphi}{\partial t}\right|_{z=-h} \approx -\left.\frac{\partial \varphi}{\partial t}\right|_{z=0} \Psi_{k_0}(-h) = \frac{g\eta}{\cosh k_0 h} = g\sqrt{1-\sigma^2}\,\eta^{(I)}, \qquad (26)$$

thus the pressure amplitude of the dominant harmonic is

$$p = g\sqrt{1-\sigma^2}\,A. \qquad (27)$$

The approximations in Eq. (26) consist in the narrowband assumption implied in the choice of the vertical mode, and also in neglecting higher-order nonlinear terms in the Bernoulli law and in the dynamics boundary condition (4). The nonlinear correction to $p_b^{(I)}$ appears in the third order of the nonlinear theory, see e.g. [Slunyaev et al., 2022].

The second harmonic pressure component $p_b^{(II)}$ was computed in [Slunyaev et al., 2022] (see their Eq. (58)); we represent it here without derivation:

$$p_2 = gk_0 \frac{(1-\sigma^2)(3-4\sigma^2)}{4\sigma^3} A^2. \qquad (28)$$

The long-scale part of the bottom pressure consists of two contributors:

$$p_b^{(0)} = \langle P_b \rangle_T = -\left.\frac{\partial \varphi^{(0)}}{\partial t}\right|_{z=-h} - \frac{1}{2}\left\langle \left(\left.\frac{\partial \varphi}{\partial x}\right|_{z=-h}\right)^2 \right\rangle_T. \qquad (29)$$

The second part of Eq. (29) gives

$$-\frac{1}{2}\left\langle \left(\left.\frac{\partial \varphi}{\partial x}\right|_{z=-h}\right)^2 \right\rangle_T \approx -\frac{1}{2(\cosh k_0 h)^2}\left\langle \left(\left.\frac{\partial \varphi}{\partial x}\right|_{z=0}\right)^2 \right\rangle_T = -\frac{\omega_0^2}{4}\frac{1-\sigma^2}{\sigma^2}|A|^2. \qquad (30)$$

If the attenuation with depth of the wave-induced long perturbation is neglected, then using Eqs. (30) and (12), the expression (29) yields

$$p_b^{(0)} \approx -\left.\frac{\partial \varphi^{(0)}}{\partial t}\right|_{z=0} - \frac{\omega_0^2}{4}\frac{1-\sigma^2}{\sigma^2}|A|^2 = g\eta^{(0)}, \qquad (31)$$

where the expression for $\eta^{(0)}$ is given by Eq. (22). In the area where the wave is absent, $A = 0$, one has $p_b^{(0)} = p_\infty = g\eta_\infty$. Note that the relation (31) which has the form of a linear hydrostatic formula, in fact takes into account second-order nonlinear corrections.

**The deep-water limit**

In the deep-water limit the expressions (22), (19), (21) and (31) give the following:

$$\eta^{(0)} \xrightarrow[k_0 h \gg 1]{} = -\frac{|A|^2}{4h} + \eta_\infty, \qquad (32)$$

$$U(x,t) \xrightarrow[k_0 h \gg 1]{} -\frac{\omega_0}{2h}|A|^2 + U_\infty = 2\omega_0 \eta^{(0)}, \tag{33}$$

$$\left.\frac{\partial \varphi^{(0)}}{\partial z}\right|_{z=0} \xrightarrow[k_0 h \gg 1]{} \frac{\omega_0}{2}\frac{\partial |A|^2}{\partial x} = -2h\omega_0 \frac{\partial \eta^{(0)}}{\partial x}, \tag{34}$$

$$p_b^{(0)} \xrightarrow[k_0 h \gg 1]{} -\frac{g}{4h}|A|^2 + p_\infty = g\eta^{(0)}, \tag{35}$$

where we have used the relation

$$\left(\frac{C_{LW}}{C_{ph}}\right)^2 = \frac{k_0 h}{\sigma}, \tag{36}$$

and the fact that in deep water $C_{ph} \approx 2C_{gr}$, see Eq. (14). The limits of solutions for the first and second harmonics (10) and (28) can be obtained straightforwardly. The limit (33) coincides with the surface condition in the theory by K. Dysthe [Dysthe, 1979].

Note that while the pressure fluctuations which correspond to the first and the second harmonics attenuate in deep water exponentially, the amplitude of the long-scale perturbation decreases inverse proportionally to the water depth, see Eq. (35). This slow dependence is a result of an implicit assumption that the modulation length exceeds the water depth, what may be expressed through the condition on the modulation wavenumber $k_{mod}$, $k_{mod}h < 1$. If the modulation length is not large enough, the regularization $C_{ph}(k) \approx C_{LW}$ in Eq. (17) and its consequences become invalid; the long-scale perturbations attenuate with depth exponentially according to the Laplace equation. Simultaneously, the attenuation of $\varphi^{(0)}$ with depth cannot be disregarded in Eq. (31), and the accurate solution yields much smaller values of the bottom pressure.

**The mean mass flux**

As is well-known, nonlinear surface waves transport mass. The normalized by the water density mean mass flux associated with the nonlinear wave propagation is the following:

$$Q(x,t) = \left\langle \int_{-h}^{\eta} \frac{\partial \varphi}{\partial x} dz \right\rangle_T = Q_U + Q_w, \tag{37}$$

$$Q_U = \left\langle \int_{-h}^{0} \frac{\partial \varphi}{\partial x} dz \right\rangle_T = \int_{-h}^{0} \frac{\partial \varphi^{(0)}}{\partial x} dz, \quad Q_w = \left\langle \int_{0}^{\eta} \frac{\partial \varphi}{\partial x} dz \right\rangle_T \approx \left\langle \eta \left.\frac{\partial \varphi}{\partial x}\right|_{z=0} \right\rangle_T = \frac{\omega_0}{2\sigma}|A|^2. \tag{38}$$

Here $Q_w$ is the Stokes drift which is calculated using the Taylor expansion (6) and taking into account only the terms of quadratic nonlinearity; $Q_U$ corresponds to the so-called return current. These functions may depend slowly on the coordinate and time. By differentiation of Eq. (37) with respect to $x$ and using the Laplace equation (1) and the bottom condition (2), we obtain

$$\frac{\partial Q}{\partial x} = \frac{\partial Q_w}{\partial x} - \left.\frac{\partial \varphi^{(0)}}{\partial z}\right|_{z=0} = -\frac{\partial \eta^{(0)}}{\partial t} = C_{gr}\frac{\partial \eta^{(0)}}{\partial x}. \tag{39}$$

The final transformations in Eq. (39) are performed using Eqs. (11) and (13). Therefore, from Eq. (39) we obtain that $Q(x) = Q_\infty + C_{gr}\eta^{(0)}(x)$, where the constant $Q_\infty$ is the mean mass flux in the area where the wave is absent. Due to the definition (37), the second-order solution reads $Q_\infty = (h + \eta_\infty)U_\infty \approx hU_\infty = C_{LW}^2\eta_\infty/C_{gr}$. Thus the excess of the mean mass flux due to the presence of wave is described by an intuitively clear relation $Q - Q_\infty = C_{gr}\eta^{(0)}(x)$. Negative values of $\eta^{(0)}$ correspond to the local mass transport in the direction opposite to the wave propagation. A non-zero value of the constant $\eta_\infty$ corresponds to the constant mass flux of the same sign in the area where the wave is formally absent.

After averaging of Eq. (37) over either a large spatial domain or large wave ensemble (which will be denoted hereafter by overlines) and assuming statistically stationary conditions, we obtain the relation for averaged quantities

$$\overline{Q} = \overline{Q}_U + \overline{Q}_w, \quad \overline{Q}_U = h\overline{U}, \quad \overline{Q}_w = \frac{\omega_0}{2\sigma}\overline{|A|^2} > 0. \tag{40}$$

If to assume that the induced flow attenuates with depth very slowly, the mean current reads

$$\overline{U} = \overline{\left.\frac{\partial\varphi^{(0)}}{\partial x}\right|_{z=0}}. \tag{41}$$

If the mean current $\overline{U}$ is zero, then the averaged mass transport $\overline{Q}$ is always positive for non-zero waves due to the inequality $\overline{Q}_w > 0$.

## 3. Generation of the wave-induced components in numerical simulations

In this section, the process of generating long-scale disturbances by surface waves is considered using the direct numerical simulation. The initial condition at $t = 0$ is taken in the form of a bell-shaped wave train with the carrier wavenumber $k_0$, placed in a large basin of the depth $h$. The initial surface displacement and the potential are related according to the linear theory. The trains are characterized by the maximum wave height of 10% of the breaking limit according to the formula suggested by Fenton [Fenton, 1990]; thus the waves are moderately nonlinear. The consequent evolution is simulated by the High Order Spectral Method [West et al., 1987] with the nonlinearity parameter $M = 3$ (up to four-wave nonlinear interactions are resolved accurately). The boundary conditions are periodic.

Evolutions of the wave trains for three cases of the water depth $k_0h = 0.4, 0.6, 1.0$ are shown in Fig. 3. The wave groups are initially located at $x = 0$ and then propagate to the right; $T_0 = 2\pi/\omega_0$ denotes the wave period. The dashed line corresponds to the rightward propagation with the group velocity $C_{gr}$. As the phase-locked bound waves start to propagate simultaneously with the 'mother' waves, but are not prescribed by the initial condition, residual waves are generated as was discussed in e.g. [Dommermuth, 2000]. The small-amplitude wave train lagging behind the main wave train is clearly seen in Fig. 3a (and is less pronounced at deeper conditions), which are waves with the dominant wavenumber $2k_0$. The solid straight blue lines indicate propagation with the long-wave velocity $C_{LW}$. Indeed, long waves following this line may be observed in Fig. 3a,b and to a less extent in Fig. 3c. The dash-dotted straight lines correspond to the leftward propagation with the velocity $C_{gr}$. Along this line one can discern some very small disturbances corresponding to waves with the dominant wave number $k_0$.

The long-wave parts of the surface perturbations obtained by using the low-pass spectral filter are shown in Fig. 4 by thick green curves. These plots confirm that together with the fast wave running with the speed $C_{LW}$ away from the main train, a long-scale set-down is generated by the main group (it propagates along the dashed straight line). The theoretical induced surface displacements are calculated using Eq. (22) with $\eta_\infty = 0$, they are shown in Fig. 4 by solid black curves. The complex wave envelope $A$ necessary for Eq. (22) was obtained from the simulated data using the band-pass spectral filter and the Hilbert transform. Its absolute value is shown in Fig. 3 by solid black curves. The theoretical result in Fig. 4 fits the numerical simulation very well as soon as discussed above secondary waves run sufficiently far away from the main train.

The long-scale velocities and dynamic bottom pressures were calculated from the simulated data in a similar way. The curves of the long-scale velocity $U$ look very similar to the solution for $\eta^{(0)}$ in Fig. 4, thus are not shown. The evolution of the long-scale bottom pressure $p^{(0)}$ is shown in Fig. 5 by green curves; it is qualitatively similar to Fig. 4. In line with radiation of fast long waves of elevation, a counter flow which is accompanied by a local depression beneath the main wave group is generated; it convoys the main train. The theoretical solution (31) taking $p_\infty = 0$ (black solid curves) agrees with the numerical simulation very well. The agreement is a bit worse for the shallowest case which may be related to the worse convergence of the wave representation in the form of a series of nonlinear wave harmonics.

## 4. Statistical moments for three wave components: General formulas

Let us now consider the effect of phase-locked bound waves in statistical sense. For this purpose we use the wave field decomposition in the form of three harmonics, similar to Eqs. (7) and (25)

$$u(\theta) = u_0(\theta) + \frac{1}{2}\left(u_1(\theta)e^{i\theta} + c.c.\right) + \frac{1}{2}\left(u_2(\theta)e^{2i\theta} + c.c.\right), \tag{42}$$

where $u_0$, $u_1$ and $u_2$ are slow functions of the phase $\theta$. The statistical moments are defined traditionally as follows: the mean $\bar{u}$, the dispersion $u_{rms}$, and the skewness $Sk^{(u)}$:

$$\bar{u} = \overline{u(x)}, \quad u_{rms}^2 = \overline{(u(x)-\bar{u})^2}, \quad Sk^{(u)} = \frac{1}{u_{rms}^3}\overline{(u(x)-\bar{u})^3}, \tag{43}$$

where the overline may be understood either as an average over a large spatial domain or as an ensemble average. By substituting Eq. (42) to (43) and assuming that oscillatory functions result in zero contributions after the averaging, by the direct calculations we obtain:

$$\bar{u} = \overline{u_0}, \quad u_{rms}^2 = \frac{1}{2}\overline{|u_1|^2} + O(\varepsilon^4), \quad Sk^{(u)} = \frac{1}{u_{rms}^3}\left[\frac{3}{2}\overline{(u_0-\bar{u})|u_1|^2} + 3\overline{\left(u_1^2 u_2^* + c.c\right)} + O(\varepsilon^6)\right], \tag{44}$$

where we have assumed that the second harmonic and the long-scale component are the second-order corrections to the first harmonic by imposing $u_1 = O(\varepsilon)$ and $u_2 = O(\varepsilon^2)$, $u_0 = O(\varepsilon^2)$ for $\varepsilon \ll 1$.

Next, we assume that the components $u_1$, $u_2$, $u_0$ are related to the complex amplitude $Y(\theta)$ through three given real coefficients $r_1$, $r_2$ and $r_0$ as follows:

$$u_1 = r_1 Y, \quad u_2 = r_2 Y^2, \quad u_0 = r_0 |Y|^2, \tag{45}$$

what is in accord with the hydrodynamic theory developed in Sec. 2. Then the formulas (44) yield

$$\bar{u} = r_0 m_2, \quad u_{rms}^2 = \frac{r_1^2}{2}m_2 + O(\varepsilon^4),$$

$$Sk^{(u)} = \frac{3m_4}{\sqrt{2}m_2^{3/2}}\left[\frac{r_2}{r_1} + 2\frac{r_0}{r_1}\left(1-\frac{m_2^2}{m_4}\right)\right] + O(\varepsilon^3) = \frac{3}{\gamma}\sigma_u\left[\frac{r_2}{r_1^2} + 2\frac{r_0}{r_1^2}(1-\gamma)\right] + O(\varepsilon^3), \quad \gamma = \frac{m_2^2}{m_4}, \tag{46}$$

where the moments of wave envelope $m_n$ have been introduced as

$$m_n = \langle |Y|^n \rangle, \quad n = 2, 4, \dots . \tag{47}$$

The moments (47) can be represented in terms of probability integrals. Under the narrowband assumption the envelope function $|Y|$ may be replaced by the wave amplitude $\zeta \geq 0$, and then the probability density function for the envelope $f(|Y|) = f(\zeta)$ means the probability density function for the wave amplitudes:

$$m_n = \int_0^\infty |Y|^n f(|Y|)d|Y| \approx \int_0^\infty \zeta^n f(\zeta)d\zeta, \quad \int_0^\infty f(\zeta)d\zeta = 1. \tag{48}$$

In particular, in the case of a uniform wave with the amplitude $\zeta = a_0$ the probability function is $f(\zeta) = \delta(\zeta - a_0)$ where $\delta(\cdot)$ is the Dirac delta function, and then $m_n = a_0^n$ and Eq. (46) yields

$$\bar{u} = r_0 a_0^2, \quad u_{rms} = \frac{r_1}{\sqrt{2}} a_0, \quad Sk^{(u)} = 3\frac{r_2}{r_1^2}\sigma_u. \tag{49}$$

For a Gaussian random process the probability density function for amplitudes $\zeta \geq 0$ is given by the Rayleigh distribution

$$f(\zeta) = \frac{\zeta}{a^2}\exp\left(-\frac{\zeta^2}{2a^2}\right), \tag{50}$$

where $a$ has the meaning of the standard deviation of the corresponding wave displacement Re $Ye^{i\theta}$, where $\zeta = |Y|$. For the Rayleigh distribution (50) $m_2 = 2a^2$ and $m_4 = 8a^4$, and then Eq. (46) yields

$$\bar{u} = 2r_0 a^2, \quad u_{rms} = r_1 a, \quad Sk^{(u)} = 6\frac{r_0 + r_2}{r_1^2}u_{rms}. \tag{51}$$

According to Eq. (46), the relative contribution of the difference harmonic to the third statistical moment is controlled by the ratio $\gamma = m_2^2/m_4$ which can vary depending on the particular wave statistics. The contribution of the difference harmonic vanishes in the case of a uniform wave ($m_2^2/m_4 = 1$), it tends to the maximum in the case of a very broad distribution of amplitudes ($m_2^2/m_4 \to 0$), simultaneously, the magnitude of skewness grows as $\gamma^{-1}$. The relative effect of the difference harmonic is intermediate for the Gaussian random process ($m_2^2/m_4 = 0.5$).

## 5. Skewness of surface displacements and bottom pressure fluctuations

Let us now discuss the skewness coefficients obtained in Sec. 4 in application to hydrodynamic fields of the surface displacement and the bottom pressure; we consider the surface displacement first. For a regular Stokes wave with the height $H$ the solution $Sk^{(\eta)}$ is given by relations (49) and (10) where $r_1 = 1$, $r_2 = (3-\sigma^2)/4\sigma^3$ and $a = k_0 H/2$. Thus in the deep-water limit $Sk^{(\eta)} = 3/8^{1/2} k_0 H/2 \approx 1.06\, k_0 H/2$, and it grows in shallower water in agreement with Fig. 1.

For random Gaussian waves the solution is provided by Eq. (51), which in terms of the surface displacement has the following form

$$Sk^{(\eta)} = 6(\alpha_0 + \alpha_2)k_0\eta_{rms}, \quad \alpha_0 = -\frac{1}{4\sigma}\frac{2C_{ph}C_{gr} + C_{LW}^2(1-\sigma^2)}{C_{LW}^2 - C_{gr}^2}, \quad \alpha_2 = \frac{3-\sigma^2}{4\sigma^3}. \quad (52)$$

It is straightforward to check that $\alpha_0 < 0$ and $\alpha_2 > 0$ at any depth, though $Sk^{(\eta)}$ remains always positive as shown in Fig. 6a (see the red curve). The individual effects of the zeroth and the second harmonics are stronger in shallow water and weaker in deep water.

The skewness when the long-scale nonlinear wave component is neglected (by imposing $\alpha_0 = 0$ in Eq. (52)) is plotted in Fig. 6a by the blue curve. In the deep water limit it reproduces the result of [Mori & Janssen, 2006] $Sk^{(\eta)} = 3k_0\eta_{rms}$, obtained within the narrowband approximation of the Zakharov equation (the situation beyond the narrowband assumption was first considered in [Longuet-Higgins, 1963], which is more complicated). We reiterate that the solution (52) overestimates the absolute value of skewness in the range of deep water.

The picture for the bottom pressure beneath random Gaussian waves is different. The solution (51) yields the following expression for the pressure skewness:

$$Sk^{(p)} = 6\frac{\beta_0 + \beta_2}{\beta_1}k_0\eta_{rms} = 6\frac{\beta_0 + \beta_2}{\beta_1^2}\frac{k_0}{g}p_{rms}, \quad (53)$$

$$\beta_0 = \alpha_0, \quad \beta_1 = \sqrt{1-\sigma^2}, \quad \beta_2 = \frac{(1-\sigma^2)(3-4\sigma^2)}{4\sigma^3}, \quad p_{rms} = g\beta_1\eta_{rms},$$

shown by the red curve in Fig. 6b. The second harmonic contributes to positive values of the third statistical moment through $\beta_2$ only in sufficiently shallow water $\sigma^2 < 3/4$, what is $k_0h < 1.317$, but is negative otherwise. This feature was emphasized in [Slunyaev et al., 2022] in relation to the problem of reconstruction of the surface displacement from the bottom pressure measurements; it was also tested there in laboratory experiments. The bottom pressure skewness solely due to the second harmonic (when $\beta_0$ is forced to zero value) is shown in Fig. 6b with the blue curve. The long-scale component is characterized by $\beta_0 < 0$ at any depth; its effect exceeds the contribution of the second harmonic and results in large negative values of the total $Sk^{(p)}$ at any depth, see the red curve in Fig. 6b. In deep water the second harmonic contributes to the negative values of $Sk^{(p)}$ as well. Note that in the limit of large depth the skewness tends to minus infinity due to the ratio $\beta_0/\beta_1 \to -\infty$. This singularity is because of the assumption that the water depth is smaller than the modulation length as discussed in the end of Sec. 2, hence it is not physical.

As discussed in Sec. 5, the relative effect on the skewness from the long-scale component increases if the wave statistics provides smaller values of $\gamma = m_2^2/m_4$; simultaneously the absolute value of skewness increases for a given variance. Therefore, whereas the situation when the long-scale effect is not taken into account at all ($\alpha = 0$ and $\beta = 0$ in Eqs. (52) and (53), see blue curves in Fig. 6) gives the upper limit of the skewness, its limit from below for certain wave amplitude distributions is absent.

To summarize, the effect of induced long-scale perturbation contributes to negative values of the wave skewness at any depth, for both the surface displacements and bottom pressure fluctuations. Due to the slow attenuation of long perturbations with depth, the relative role of the long-scale part is stronger in the case of the bottom pressure field; then the theoretical skewness of Gaussian waves is negative at any depth. Specific wave distribution may lead to further decrease of the skewness such that it can become negative for the surface displacement too.

Let us now compare the theoretical predictions with the results of numerical simulations presented in Sec. 3. The evolution of the skewness coefficients $Sk^{(\eta)}$ and $Sk^{(p)}$ for the simulated at different depths wave trains are shown in Fig. 7a and Fig. 7b relatively by symbols. The corresponding theoretical values calculated according to Eq. (46) are plotted by solid curves; the moments $m_2$ and $m_4$ are calculated from the obtained complex envelope functions $A$. The theoretical solution fits the simulated data well when the main train is cleared of waves of other lengths. There is systematic underestimation by absolute value of the skewness of simulated surface displacement, which becomes more significant in shallower water. In case of the bottom pressure the agreement between the theory and simulations is excellent. The instantaneous skewness coefficients decrease by absolute values over time due to the continuous spreading of waves caused by the dispersion.

Note that in all cases shown in Fig. 7 the skewness coefficients become negative shortly after the start of simulations. This fact follows from the direct computation of the skewness and from the analytic solution based on the complex amplitude of the envelope, both. To understand this surprising result, let us examine a piece of the simulated surface shown in Fig. 8, which corresponds to a late stage of the evolution for the case $k_0 h = 0.6$. The skewness coefficient $Sk^{(\eta)}$ is calculated for three intervals of the displayed surface: for the entire sample, for the wave group excluding the leading long wave (pink selection) and for the section which corresponds to relatively uniform waves (shown in green); the corresponding values are given in the legend. In agreement with the plots in Fig. 7a, the entire series is characterized by significant negative value of the skewness, $Sk^{(\eta)} = -0.22$. Removing the long wave from the series reduces the mean elevation, and thus the function $\eta - \bar{\eta}$ for the pink interval is effectively shifted upward compared to the full sample. Consequently, the value of $Sk^{(\eta)}$ increases to $-0.06$. The negative value of skewness is obviously due to the particular distribution of wave amplitudes which differs from the Rayleigh law. When the central part of the wave group is considered, the induced set-down is almost constant and thus is effectively absent in the series $\eta - \bar{\eta}$ used for the calculation of the statistical moment. Thus, the reduction of skewness due to the effect of long-scale depression is weakened, and the main contribution to the skewness comes from the second harmonic, which becomes positive as should be for Stokes waves. According to Eq. (49), for the depth $k_0 h = 0.6$ the skewness of a regular wave is $Sk^{(\eta)} \approx 9.3 k_0 a_0$ where $a_0$ is the wave amplitude, what is close to the directly calculated value.

## 6. Models for wave group compositions

It is clear that the negative skewness values for any depth predicted by the solution (53) (the red curve in Fig. 6b) do not correspond to the picture of in-situ measurements shown in Fig. 2. Let us discuss possible reasons of this discrepancy.

Averaging of the condition (15) over a large area or a large ensemble gives (the carrier wavenumber $k_0$ is assumed to be constant)

$$g\bar{\eta} - C_{gr}\bar{U} = -\frac{\omega_0^2}{4}\frac{1-\sigma^2}{\sigma^2}\overline{|A|^2}, \quad \bar{\eta} = \overline{\eta^{(0)}}, \quad \bar{U} = \overline{\frac{\partial \varphi^{(0)}}{\partial x}}\bigg|_{z=0}. \tag{54}$$

For slowly attenuating with depth long-scale perturbations the quantity $\bar{U}$ has the meaning of a uniform background current. From Eq. (54) it inevitably follows that if the water is not infinitely deep, then any waves propagate with non-zero either mean displacement $\bar{\eta}$ or mean current $\bar{U}$ or both. Their expressions, and also the mean pressure $\bar{p}$ appear as results of averaging of Eqs. (22), (19) and (31):

$$\bar{\eta} = -\frac{k_0}{4\sigma}\frac{2C_{ph}C_{gr} + C_{LW}^2(1-\sigma^2)}{C_{LW}^2 - C_{gr}^2}\overline{|A|^2} + \bar{\eta}_\infty, \tag{55}$$

$$\overline{U} = \frac{-gk_0}{4\sigma} \frac{2C_{ph}C_{gr} + C_{gr}^2(1-\sigma^2)}{C_{gr}(C_{LW}^2 - C_{gr}^2)} \overline{|A|^2} + \overline{U}_\infty, \qquad (56)$$

$$\overline{p} = g\overline{\eta}. \qquad (57)$$

If the wave perturbations are localized within some spatial interval and do not disturb water far from them (i.e., embedded in still water), the constants $\overline{\eta}_\infty$ and $\overline{U}_\infty$ vanish, and then $\overline{U}$ and $\overline{\eta}$ are both not zeros.

The set-downs are clearly seen in Fig. 4; the return flows $U(x,t)$ beneath wave groups look very similar to these graphs. In the numerical simulations the mean elevation and mean velocity are zeros over the simulation time due to the conservation of volume and momentum by the hydrodynamics equations. The inconsistency between the linear initial condition and the nonlinear structure of travelling waves is resolved through the generation of new waves. The emerging long wave of elevation carries mass to the right direction. Respectively, the main wave train should carry similar mass to the left. As discussed just after Eq. (39), the mean mass flux due to the presence of waves is proportional to $\eta^{(0)}(x)$, and hence it is indeed negative for the main train. The formula for the excess of the averaged mass flux due to the presence of waves follows from Eq. (39):

$$\overline{Q} - \overline{Q}_\infty = C_{gr}\overline{\eta}. \qquad (58)$$

In the limit of deep water $\eta^{(0)} = \eta_\infty$ (see Eq. (32)), and if the reference is $\eta_\infty = 0$, then $\overline{\eta} = 0$ and the positive Stokes drift $\overline{Q}_w$ is compensated by the negative return flow $\overline{Q}_U$ as was discussed in the literature (e.g. [Fabrikant & Stepanyants, 1998]) (note that the relation $\overline{Q}_U = h\overline{U}$ fails in the limit of very deep water). It is clear that a statistical ensemble of groups like those shown in Fig. 3 will possess a negative mean displacement $\overline{\eta}$, negative mean current $\overline{U}$ and consequently negative mass flux, what looks questionable from the physical point of view.

We reiterate that the joint condition $\overline{U} = 0$ and $\overline{\eta} = 0$ cannot be fulfilled in finite-depth basins. The regular Stokes wave solution is characterized by zero background flow, and then from Eq. (54):

$$\overline{\eta} = -\frac{k_0}{4}\frac{1-\sigma^2}{\sigma}\overline{|A|^2} < 0, \quad \overline{U} = 0, \quad \overline{\eta}_\infty > 0, \quad \overline{U}_\infty > 0. \qquad (59)$$

Under the assumption $\overline{U} = 0$ the averaged linear part of the Bernoulli law (31) returns zero, the nonlinear part of the Bernoulli law yields exactly the relation (57). Comparing Eqs. (59) and (55), it becomes clear that the effects of induced set-down and bottom pressure reduction are weakened under the assumption of a zero mean current. The set-down in the form (59) appears not only in application to uniform Stokes waves (see Eq. (3.6) in [Zhao & Liu, 2022]), but also was obtained for nonlinear-dispersive waves in [Brinch-Nielsen & Jonsson, 1986] and in the theory of shoaling nonlinear waves without dissipation within the framework of the energy balance equation (see Eq. (7.4.22) in [Holthuijsen, 2007]).

It is easy to see that a non-zero value of $\eta_\infty$ (and respectively $U_\infty$) will not influence eventual values of the central statistical moments (Eq. (43)) if it is assumed to be universal for all groups constituting the wave ensemble: $\overline{\eta}_\infty = \eta_\infty$. Let us assume that for *each nominal group* the constants $\eta_\infty$ and $U_\infty$ in the solutions (19) and (22) are chosen to ensure zero mean current, $<U>_G = 0$. This means that the parameters $\eta_\infty$ and $U_\infty$ are not universal constants but are functions of intensity $<|A|^2>_G$ characterizing particular wavegroup. Hereafter $<\cdot>_G$ denotes averaging over the group size, which is greater than the wave length. The large-ensemble averaging will be assumed a two stage procedure with the averaging over the group scale at first, and then averaging over a large number of groups. Naturally, the averaged mean current will be zero, $\overline{U} = 0$, and $\overline{\eta} < 0$ according to Eq. (54). In this case the consideration in Sec. 4 remains valid, but the coefficients of relations between the difference harmonic and the wave envelope ($r_0$ in Eq. (45) and, correspondingly, $\alpha_0$ in Eq. (52) and $\beta_0$ in Eq. (53)) become modified according to Eq. (59):

$$\widetilde{\alpha}_0 = \widetilde{\beta}_0 = -\frac{1-\sigma^2}{4\sigma}, \tag{60}$$

what for the range of intermediate depths $0.4 < k_0 h < 2$ corresponds to a reduction by an order of magnitude or more, see Fig. 9. The modification of coefficients $\alpha_0 = \widetilde{\alpha}_0$ and $\beta_0 = \widetilde{\beta}_0$ results in significant change of the skewness coefficients shown by black dashed curves in Fig. 6. The modified skewness coefficients are rather close to the ones calculated for the second harmonic only (when $\alpha_0 = 0$ and $\beta_0 = 0$) shown in Fig. 6 by blue curves. The surface displacement skewness is always above $3k_0\eta_{rms}$; the bottom pressure skewness becomes negative when $\sigma^2 < 3/5$, what is when the depth exceeds the value of $k_0 h \approx 1.032$.

A group which does not disturb the flow far from it ($\eta_\infty = 0$) and possesses non-zero setdown and negative return flow is shown schematically in Fig. 10a; hereafter it will be referred to the group model A. The second alternative with the zero mean current $<U>_G = 0$ but positive values of $\eta_\infty$ and $U_\infty$ is shown in Fig. 10b (model B). The excess of the averaged mass flux $\overline{Q} - \overline{Q}_\infty$ in groups B is smaller than in groups A. The long-scale displacement and long-scale pressure calculated using the reduced values of coefficients (60) are plotted in Fig. 4 and Fig. 5 by red curves. It is clear that they strongly underestimate the simulation results.

The third natural option is to require that the mean elevation is zero, $\overline{\eta} = 0$, and then

$$\overline{\eta} = 0, \quad \overline{U} = \frac{\omega_0^2}{4 C_{gr}} \frac{1-\sigma^2}{\sigma^2} \overline{|A|^2} > 0, \quad \overline{\eta}_\infty > 0, \quad \overline{U}_\infty > 0, \tag{61}$$

see the sketch of a single group in Fig. 10c (model C). According to Eq. (58), this wave group configuration ensures zero mass transport. Under this assumption the long-scale pressure $\overline{p}$ is also zero due to the general relation (31), and thus in the statistical formulas (52) and (53) one should put $\alpha_0 = 0$ and $\beta_0 = 0$. Hence the skewness is affected only by the second harmonic as shown by blue curves in Fig. 6. This solution does not describe the long-scale induced wave components of the surface displacement and the pressure shown in Figs. 4–5 at all.

Thus the three basic alternative wave group configurations displayed in Fig. 10 correspond to remarkably different effects of the induced long-scale perturbations on the skewness, especially when the bottom pressure is concerned. The characteristic features of these group models and the corresponding values of the coefficients $\alpha_0 = \beta_0$ are summarized in Table 2.

Let us estimate the value of $\alpha_0 = \beta_0$ based on the in-situ data shown in Fig. 2 assuming that the waves obey the Gaussian statistics. First, we involve the natural assumption that the skewness is proportional to the wave intensity, while the latter for the bottom pressure may be expressed through the dimensionless root-mean-square pressure $k_0 p_{rms}/g$ (see Eq. (53)). The ratio $gSk^{(p)}/(k_0 p_{rms})$ calculated for the instrumental data is presented in Fig. 11a by gray dots and isolevels for the data density. The scatter of data is large, but is approximately symmetrical with respect to the horizontal line. Locations of the curves which correspond to the obtained solutions for $Sk^{(p)}$ (53) in the axes of Fig. 11a do not depend on the wave intensity. They are shown by different curves for the three wave group configurations. The curves for the groups of the type A and C envelope the main amount of the data (63%); the case B is very close to the model C.

Next, we invert the formula (53) as follows:

$$\beta_0^{(instr)} = Sk^{(p)} \frac{g\beta_1^2}{6k_0 p_{rms}} \frac{\gamma}{1-\gamma} - \frac{\beta_2}{2(1-\gamma)}, \qquad (62)$$

where the coefficients $\beta_1$ and $\beta_2$ are still defined as per Eq. (53), $\gamma = m_2^2/m_4 = 0.5$, and $k_0 p_{rms}$ and $Sk^{(p)}$ come from the instrumental data. The result is shown in Fig. 11b by gray dots and density isolevels. The curves corresponding to solutions for the three wave group configurations are also plotted; the type C corresponds to the zero value of $\beta_0$. Similar to Fig. 11a, we observe in Fig. 11b that most of the data is located between the solutions for the groups of the type A (groups are embedded in still water) and type B (zero mean current). Two empirical curves are plotted in Fig. 11b in addition:

$$\beta_0^{(fit1)} = 4\widetilde{\beta}_0 \quad \text{and} \quad \beta_0^{(fit2)} = -\frac{1}{2(k_0 h)^3}. \qquad (63)$$

They both seem to describe the instrumental data reasonably well, and are close to each other in the interval $k_0 h > 0.8$. The dependence $(k_0 h)^{-3}$ naturally appears in the shallow water limit of the coefficient $\alpha_0$, though with a different coefficient: $\alpha_0 \approx -3/4(k_0 h)^{-3}$ for $k_0 h \ll 1$. At the same time, the scatter of the instrumental data is huge leading to small formal characteristics of coincidence. In particular the coefficient of determination $R^2$ is of the order of 10%.

Finally, symbols in Fig. 11b represent the result of numerical simulations of irregular directional water waves within the Euler equations (to be reported in detail elsewhere; similar simulations of the surface displacement but for deep water were presented in [Slunyaev & Kokorina, 2023]). The simulations were performed for the conditions of constant water depth in domains of the size about 50 by 50 wave lengths. The initial conditions were specified according to the realistic for sea waves JONSWAP spectrum with random wave phases and given characteristic wave intensities and directional spreads according to the $\cos^2$ distribution. To reduce the amount of spurious waves emerged due to the inconsistency between the linear initial condition and nonlinear equations, each simulation started with a nonlinear adaptation stage, when the nonlinear terms of the equations were slowly brought into play following the approach by [Dommermuth, 2000].

Symbols in Fig. 11b correspond to the result of the bottom pressure processing in four cases of depths, two cases of widths of the angular spectra and several weak to moderate intensity cases (before the wave breaking effect becomes significant). In the figure blue circles correspond to long-crested waves with the spread 12°, while red crosses indicate the cases of broader spectrum with the maximum spread 62°. For the numerical simulations, the ratio $\gamma = m_2^2/m_4$ is computed for every set of conditions. It is found to be within the interval from 0.47 and 0.63, and is taken into account when computing Eq. (62). Larger values of $\gamma$ occur under conditions of shallower water, little directional spread and high wave intensity. One can see that the points exhibit very consistent results. It is noteworthy that for $k_0 h \geq 0.6$ it is impossible to distinguish on the graph situations corresponding to different wave intensities: they give very close estimates of $\beta_0$. One can notice a very good agreement between the data of numerical simulations and the most probable values of $\beta_0$ following from the in-situ measurements. Again, the simulated data is located between the curves which correspond to the group models A and B. Waves with narrower angular spectra (blue circles) are closer to the model A. In the shallowest case $k_0 h = 0.4$ the results of numerical simulations of waves of different intensities agree slightly worse, but are very well approximated by the second fitting curve given in Eq. (63).

## 7. Conclusions

It is conventionally assumed that the water wave positive skewness is produced by the second nonlinear harmonic which alters the shape of individual waves. In this work we emphasize the importance of the effect of induced long-scale perturbations (return flow and group set-down) on the coefficient of skewness. It becomes most dramatic in the case of the pressure field beneath waves in intermediate depth due to the slow attenuation of the long perturbations with depth, and may result in large negative values of the skewness. Validity of the developed theory requires that the characteristic length of wave modulations is sufficiently large compared to the local depth. If this condition breaks, the low frequency component of the bottom pressure decreases, and its effect on the skewness becomes negligible.

Depending on the configuration of groups representing the stochastic wave field, the influence of the induced long-scale perturbations may have remarkably different strengths. We have suggested and considered three basic group structures displayed in Fig. 10, which represent groups embedded in still water (model A), groups with zero mean current (model B) and groups with zero mean displacement (model C). Groups of this or that type should prevail under specific physical conditions.

We have shown that groups of the first type (model A) appear in the initial-value problem with linear initial conditions. Among the three types of the groups this case corresponds to the strongest effect of the long-scale perturbation leading to significant decrease of the surface displacement skewness and negative values of the bottom pressure skewness of random waves. The generation of non-zero mean wave displacement results in the difference between the mean water depth and still water depth, see the review in [Zhao & Liu, 2022]. Travelling waves with non-zero mean displacement perform mass transport.

An ensemble of wave groups characterized by zero mean currents (model B) exhibits greatly reduced effect of the wave-induced flow, which has little influence on the surface displacement skewness which is positive at any depth for Gaussian random waves, but is important for the bottom pressure skewness. Based on this model, bottom pressure records beneath irregular Gaussian waves should be characterized by negative skewness if $k_0h > 1.032$. The zero mean current conditions should be natural for numerical wave tanks which use periodic boundary conditions for the velocity potential.

If wave groups have zero mean displacement (model C), then within the second-order nonlinear theory the skewness is controlled solely by the second harmonic, thus the second-order regular Stokes wave solution may be used. At any depth the surface displacement possesses positive skewness, while the bottom pressure skewness is positive for $k_0h < 1.317$ and is negative otherwise. The zero mean displacement wave group configuration should be more likely to occur in closed basins such as experimental tanks or near-shore regions.

The direct comparison of the theory with in-situ measurements of the bottom pressure in the coastal zone and with results of stochastic numerical simulations reveals that the vast majority of the field data and also data from the numerical simulation correspond to the situation between the models A and B. The bottom pressure skewness is positive in shallow water and is negative in sufficiently deep water. The in-situ data is much scattered thus it may represent a combination of different wave group types and/or be influenced by unaccounted for physical effects. The latter could be, for example, shear currents in water; the return flow may be greatly modified by the water stratification which will couple it to internal gravity waves [McIntyre, 1981]. This list may be continued. However, the good agreement of the in-situ data with results of the numerical stochastic simulations rather indicates that the unaccounted effects lead to the data scatter but do not change the situation in essence.

The assumption of a narrow spectrum is an obvious weak point of the developed theory. It may be improved by taking into account dispersive terms. However, then the solution will require more information about irregular waves.

Some possible reasons of the in-situ data large spread become apparent from the present research. The effect of induced long-scale perturbations can be significantly enhanced for certain wave distributions. The skewness coefficient can change significantly when nonlinear groups are mixed with long waves. Long waves of elevation effectively decrease the total skewness of the record, while long waves of depression should produce an opposite effect. If the long waves are random, their net effect should vanish after averaging over a large wave ensemble. However, in conditions of relatively shallow depth, one can expect a more probable occurrence of soliton-type waves of elevation. Meanwhile, since relatively short wave samples are typically used for oceanographic purposes, the sampling variability effect may be particularly strong in the mixed wave conditions.

The obtained theoretical results can be directly verified using the data of simultaneous registration of the surface displacement and the near-bottom pressure or fluid velocities. The results can be useful in solving the inverse problem, when information about surface waves is reconstructed based on bottom pressure measurements. Particularly, as the bottom pressure skewness is shown to be sensitive with respect to the wave pattern configuration and related mean currents and mass transport induced by surface waves, it may be used for indirect assessment of these quantities. According to the presented findings, the distribution of measured bottom pressure fluctuations characterized by positive skewness may indicate stable coastal configurations, while significant negative skewness should indicate an ongoing process of intense erosion or accretion of the coast. These problems are crucial for coastal engineering, but are beyond the scope of the present study and may be a subject of further research.

## Acknowledgements

The authors are grateful to Prof. E.N. Pelinovsky and Prof. V.I. Shrira for useful discussion of the work. The theoretical study is performed for the project RSF 24-47-02007. The in-situ data processing is supported by the grant RSF 22-17-00153.

## Data availability

The data of numerical simulations that support the findings of this study are available from the corresponding author upon reasonable request.

Table 1. Key notations

| | |
|---|---|
| $h$ | position of the water bottom, $z = -h$ |
| $\sigma$ | hyperbolic tangent of the dimensionless depth $kh$ |
| $C_{ph}$ | phase velocity |
| $C_{gr}$ | group velocity |
| $C_{LW}$ | long-wave velocity |
| $\eta(x,t)$ | surface displacement with respect to the rest level $z = 0$ |
| $\varphi(x,z,t)$ | velocity potential in the volume |
| $P(x,z,t)$ | dynamic pressure |
| $A(x,t)$ | complex envelope for the surface displacement |
| $B(x,t)$ | complex envelope for the surface velocity potential |
| $p(x,t)$ | complex envelope for the dynamic bottom pressure |
| $U(x,t)$ | long-scale horizontal velocity near the surface |
| $Q(x,t)$ | horizontal mass flux |
| $\alpha_0, \alpha_1, \alpha_2$ | coefficients of the surface displacement harmonics |
| $\beta_0, \beta_1, \beta_2$ | coefficients of the bottom pressure harmonics |
| $\langle\cdot\rangle_G, \langle\cdot\rangle$ | group averaging and large wave ensemble averaging respectively |

Table 2. Wave group models

| | A | B | C |
|---|---|---|---|
| Characteristic feature | embedded in still water, $\eta_\infty = 0$ | zero mean current, $\langle U\rangle_G = 0$ | zero mean set-down, $\langle\eta\rangle_G = 0$ |
| Coefficients $\alpha_0 = \beta_0$ | $\alpha_0$, Eq. (52) | $\tilde{\alpha}_0$, Eq. (60) | 0 |

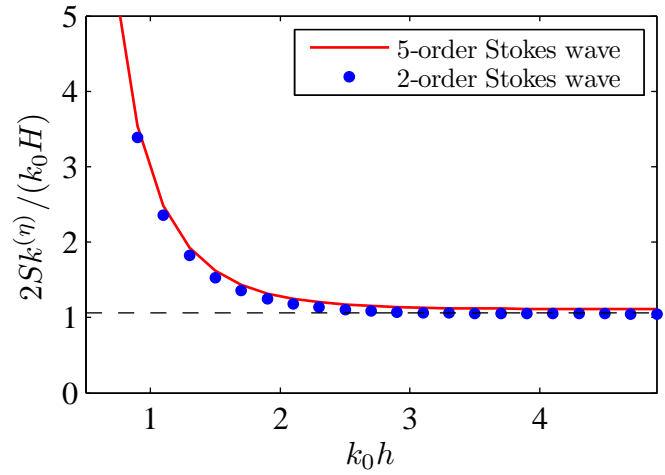

Fig. 1. Skewness $Sk^{(\eta)}$ over the steepness $k_0H/2$ for Stokes waves with wavenumber $k_0$ and crest-to-trough height $H$ in different depths $h$. The wave height is taken equal to 40% of the breaking limit according to [Fenton, 1990].

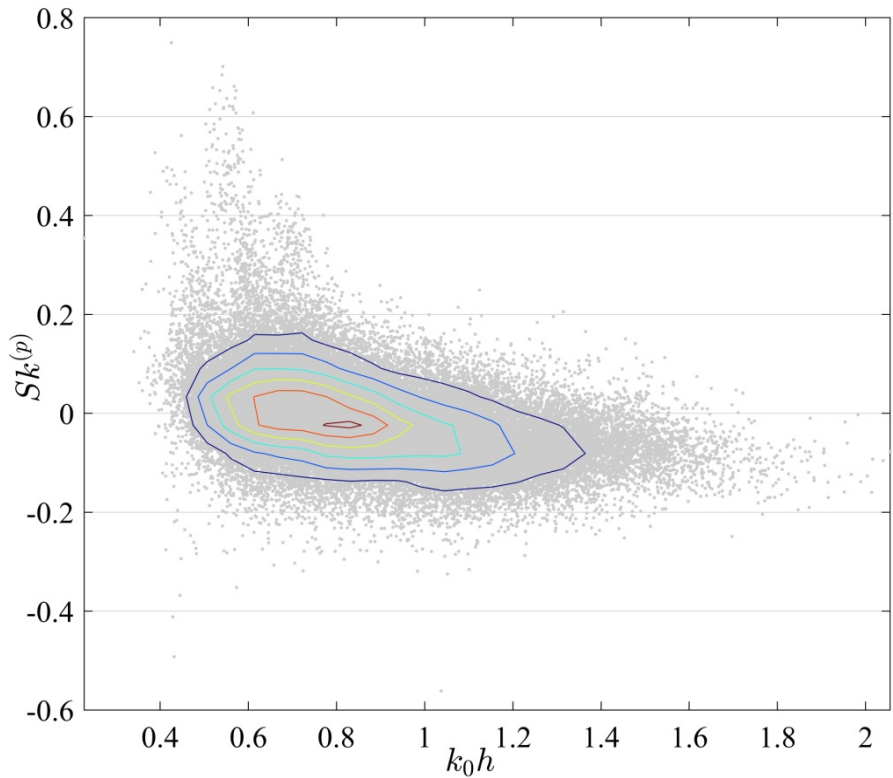

Fig. 2. Skewness $Sk^{(p)}$ of the bottom pressure records near Sakhalin Island versus the dimensionless depth parameter $k_0h$. Isolines characterize the density of in-situ data.

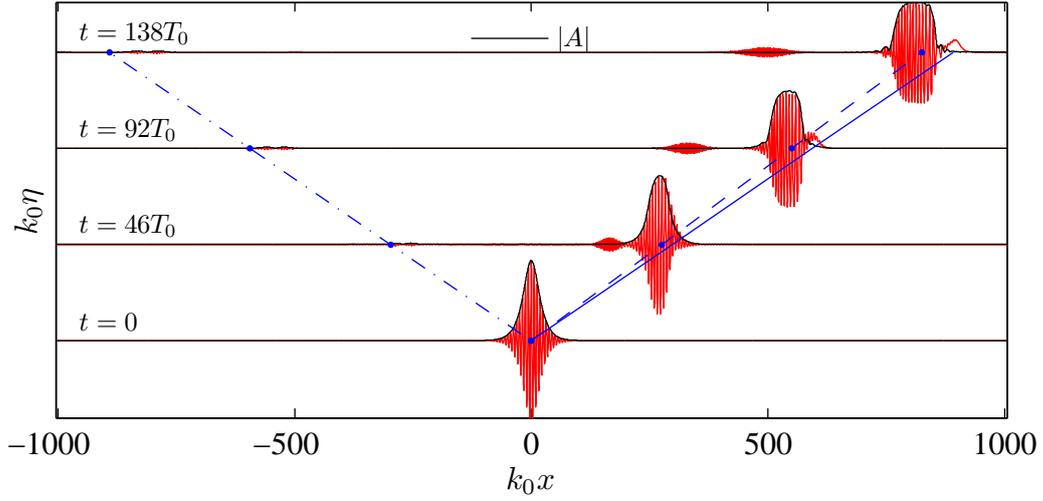

(a)

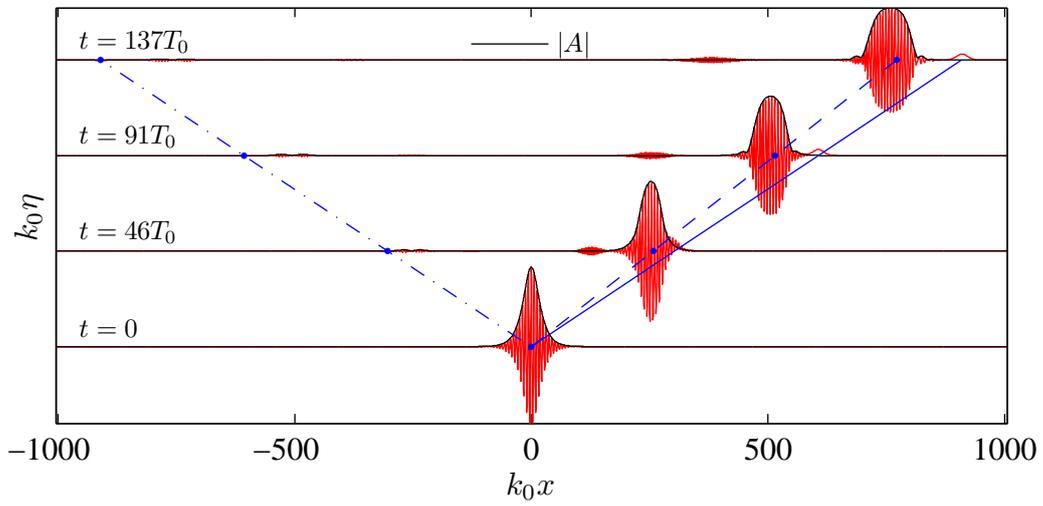

(b)

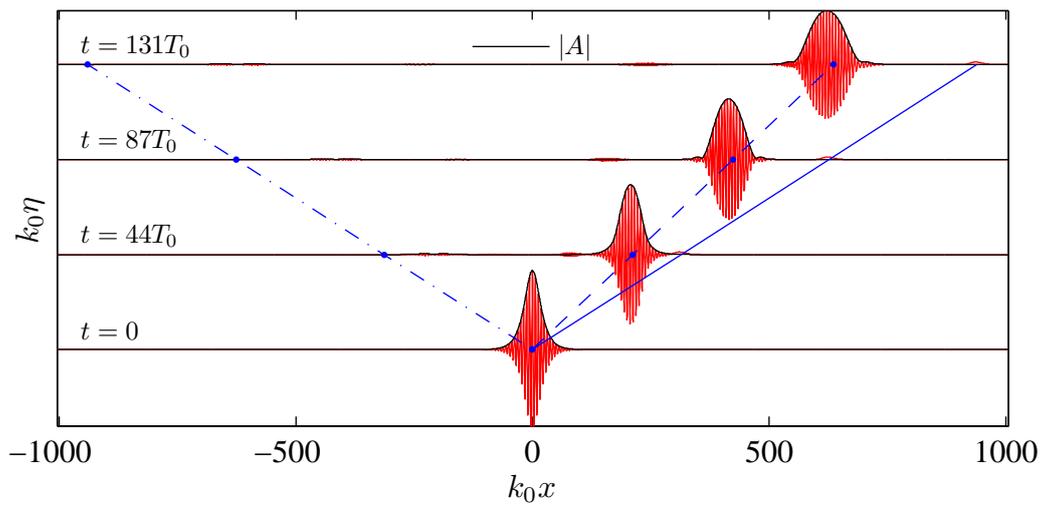

(c)

Fig. 3. Numerical simulations of the evolution of surface displacements (red curves) for the depth conditions $k_0 h = 0.4$ (a), $k_0 h = 0.6$ (b) and $k_0 h = 1.0$ (c). The solid black curve corresponds to the evaluated wave envelope $|A|$. The dash-dotted, dashed and solid straight blue lines correspond to the propagation with velocities $-C_{gr}$, $+C_{gr}$ and $+C_{LW}$ respectively.

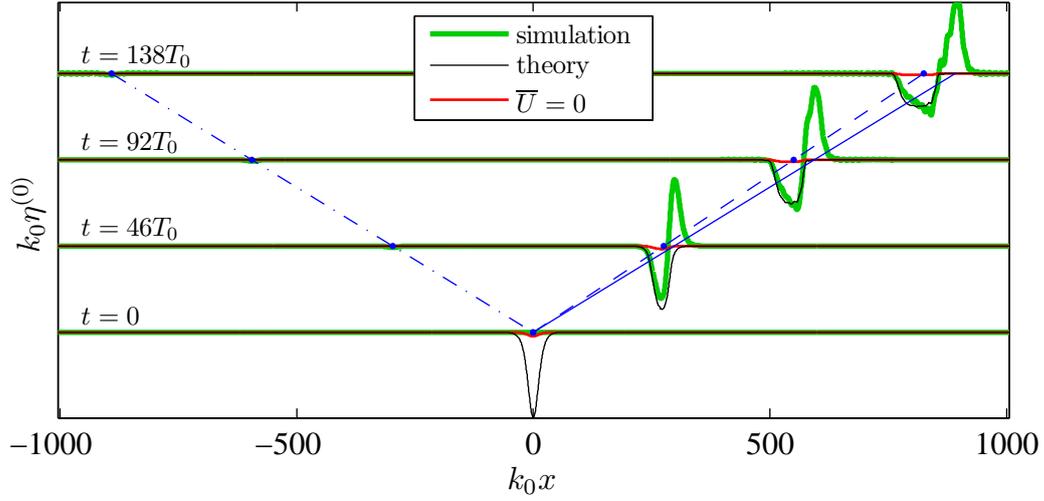

(a)

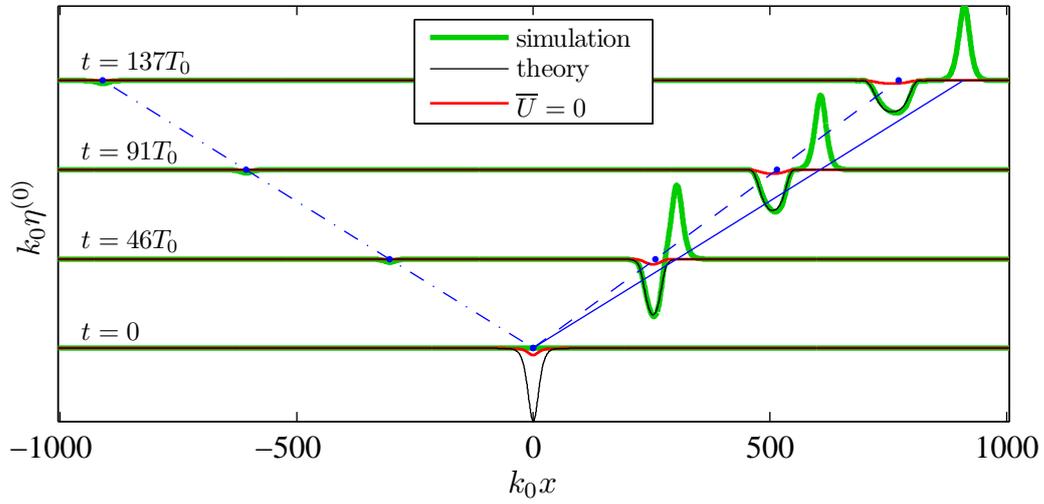

(b)

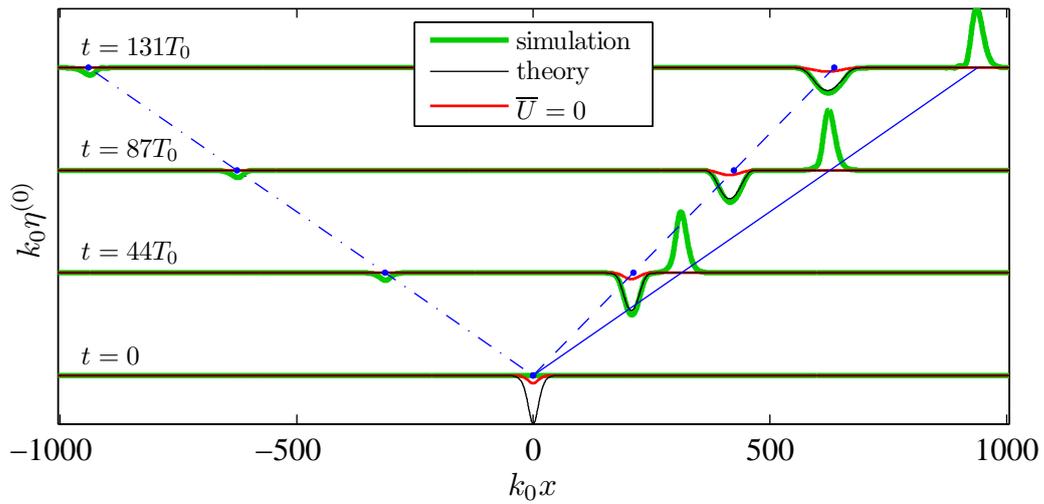

(c)

Fig. 4. Same as in Fig. 3 but the long-scale part of the surface displacements is shown (thick green curves). The black curves correspond to the solution (22) with $\eta_\infty = 0$; the red curves show the reduced effect according to Eq. (59).

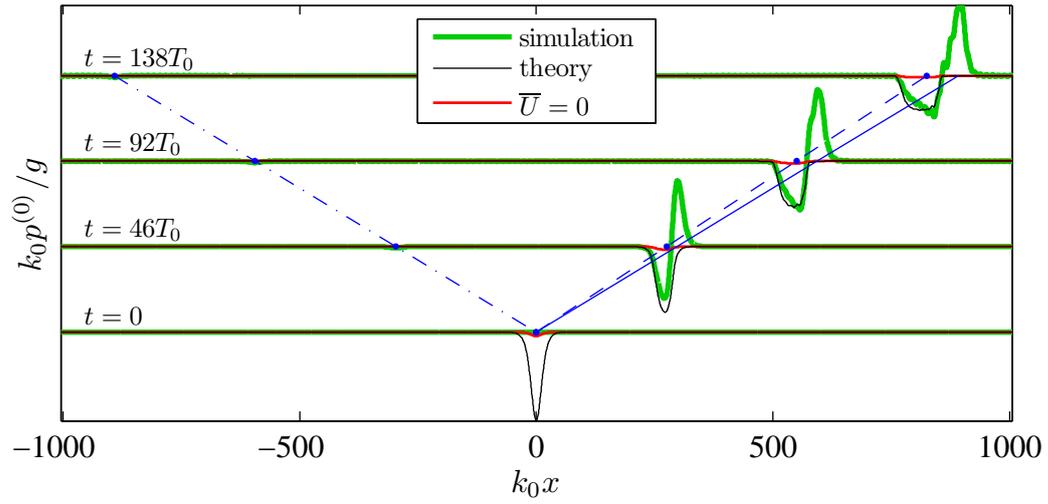

(a)

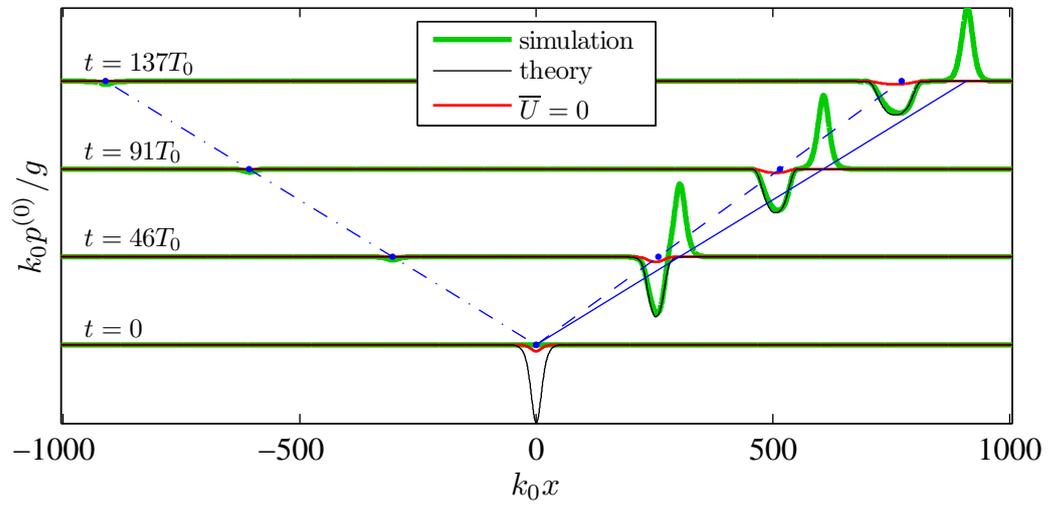

(b)

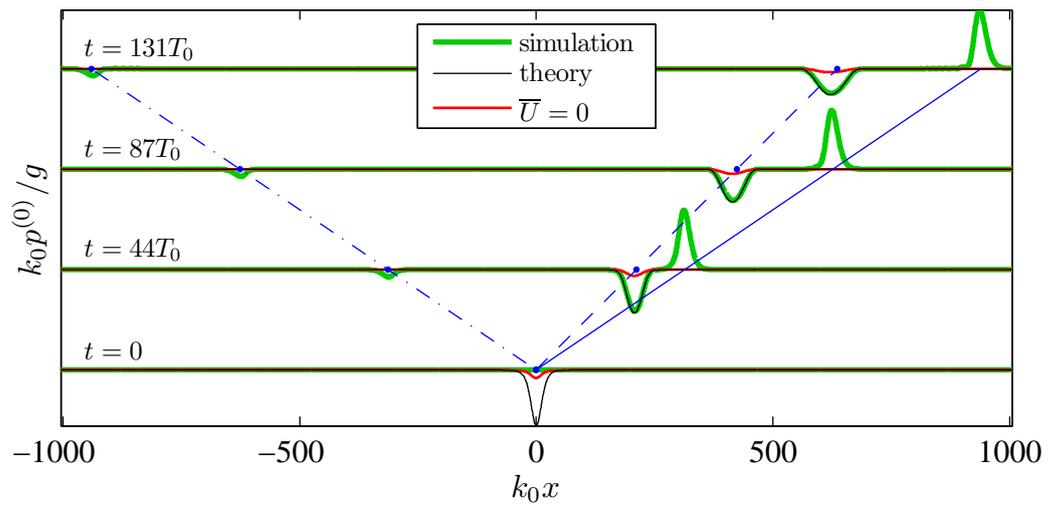

(c)

Fig. 5. Similar to Fig. 3–4 but the long-scale part of the bottom dynamic pressure is shown.

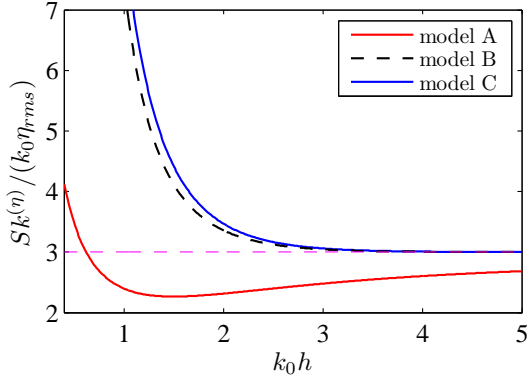 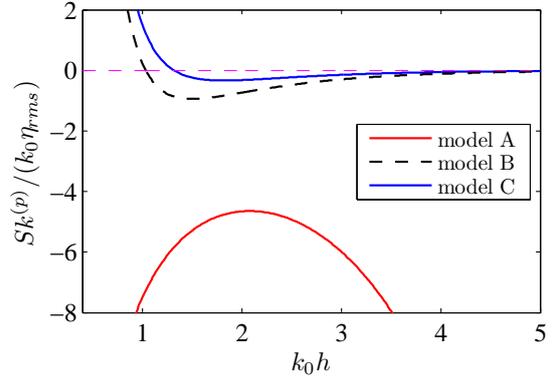

Fig. 6. Coefficients of skewness for the surface displacement (a) and the bottom dynamic pressure (b) for random waves according to Eqs. (52) and (53) (red curves), and also when $\alpha_0 = 0$ and $\beta_0 = 0$ (blue curve) and when the modified coefficients (59) are used (black dashed curves).

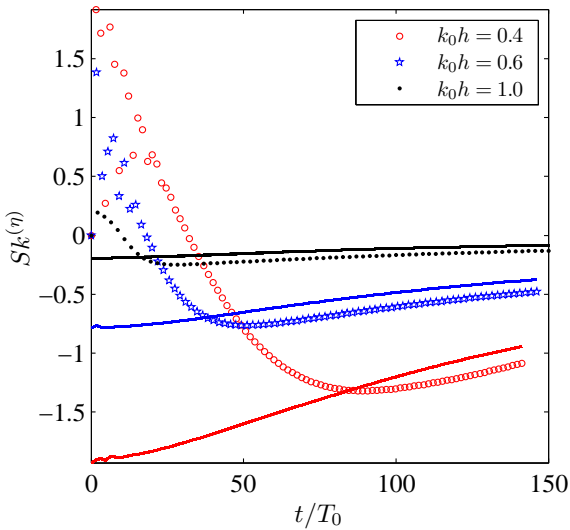 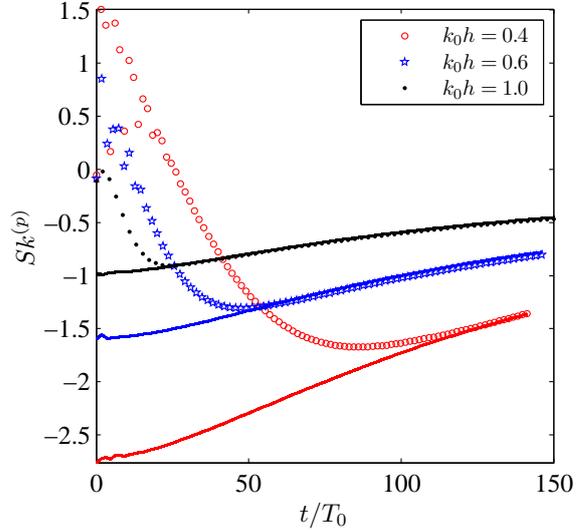

Fig. 7. Skewness of the surface displacements $Sk^{(\eta)}$ (a) and of the bottom pressure fluctuations $Sk^{(p)}$ (b) in the numerical experiments shown in Figs. 3–5. Symbols correspond to the directly computed skewness as per Eq. (43); the solid curves correspond to the solution (46) where the moments $m_2$ and $m_4$ are calculated for the envelope $A$.

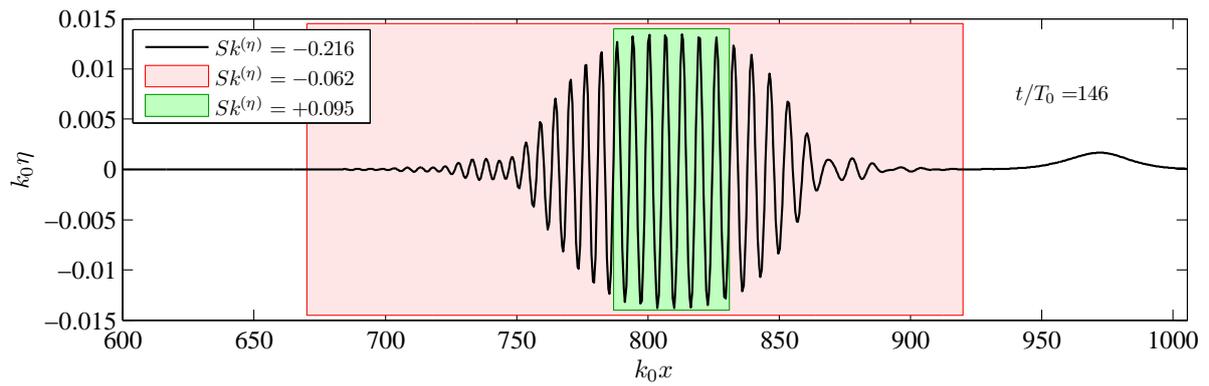

Fig. 8. A part of the simulated water surface for the depth $k_0h = 0.6$ at $t = 146\ T_0$ (see Fig. 3b). The coefficients of skewness $Sk^{(\eta)}$ calculated for different sections of the series are given in the legend.

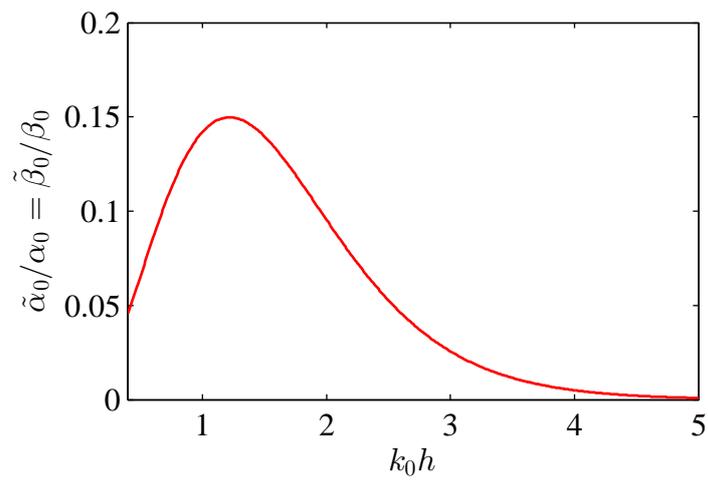

Fig. 9. Ratio of the coefficients for the long-scale induced wave component.

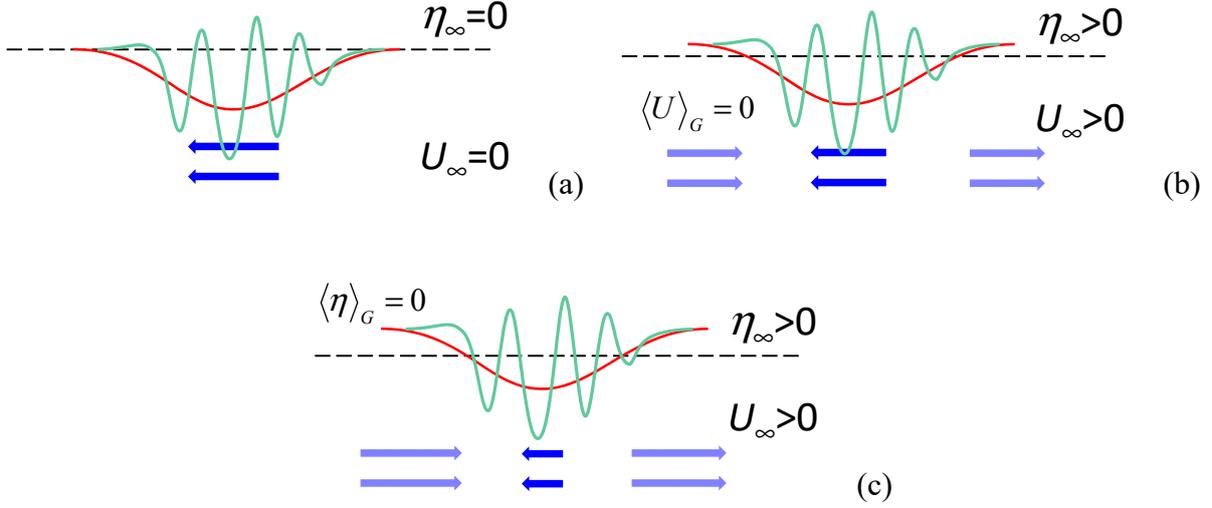

Fig. 10. Three configurations of the wave group: model A (a), model B (b) and model C (c). The long-scale displacement $\eta^{(0)}$ is shown with the red curve. The flow beneath the group is shown with blue arrows.

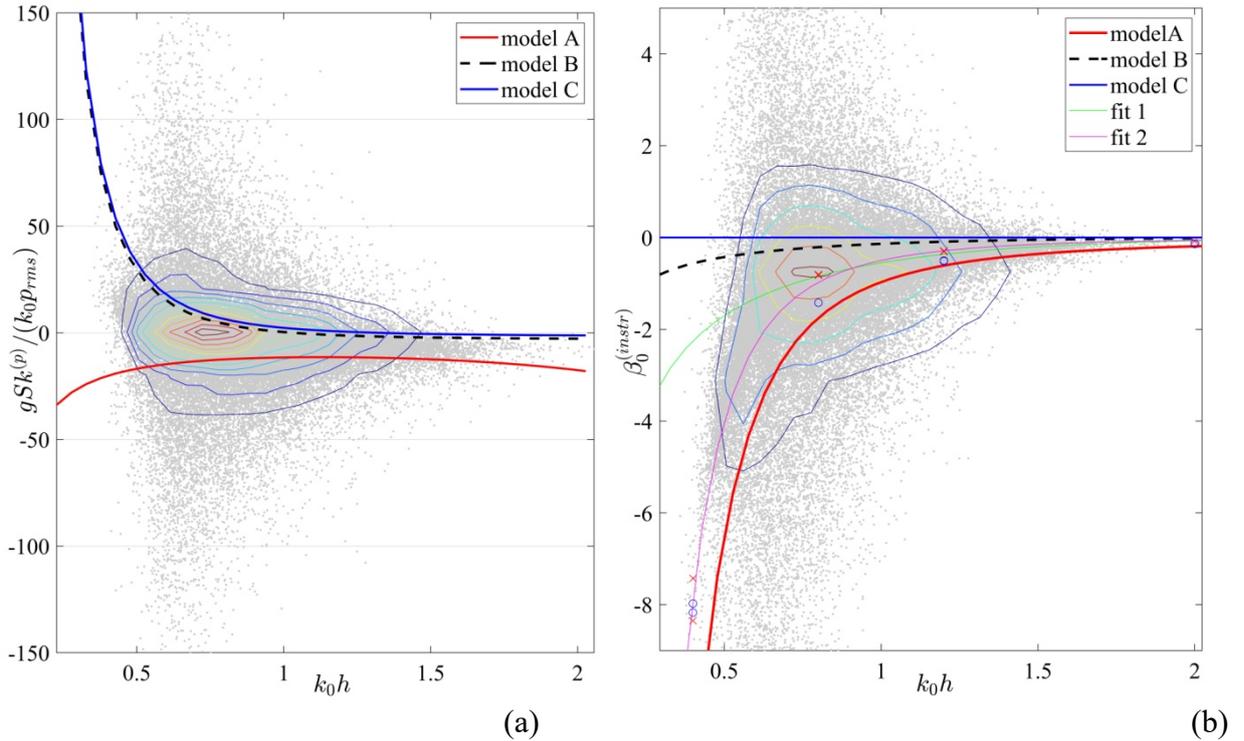

Fig. 11. Comparison of the in-situ data of the bottom pressure with the theory. (a): The skewness over standard deviation value as function of dimensionless depth $k_0 h$. (b) Estimation of the coefficient $\beta_0$ according to Eq. (62). Lines correspond to analytic solutions and fits according to formulas (63), see the legend. Symbols represent the result of the direct numerical simulations of large ensembles of waves.